\documentstyle[12pt]{article}

\begin{document}

\title{Pion form factor at very large $Q^2$}
\author{Bing An Li\\
Department of Physics and Astronomy, University of Kentucky\\
Lexington, KY 40506, USA}

\maketitle

\begin{abstract}
The pion form factor at very large $Q^2$ is obtained by using a new pion wave function. It shows that 
when $Q^2>>(1.8GeV)^2$ the pion form factor reaches the asymptotic limit ${\alpha_s(Q^2)\over Q^2}$.
\end{abstract}
\newpage
Pion form factor is a very important quantity in hadron physics.
The current matrix element of pion is written as 
\begin{equation}
<\pi^+|j_\mu(0)|\pi^+>=\int d^4 k_2\int d^4 k_1 Tr\{\phi_\pi(k_1,p_f)T_H(k_1,k_2,p_f,p_i)_\mu
\phi_\pi(k_2,p_i)\}=F_\pi(Q^2)P_\mu,
\end{equation}
where $\phi_\pi$ is the wave function of pion, $k_1,k_2$ are internal momenta of pion, $p_i$ and $p_f$
are the momenta of initial and final pion respectively, $T_H$ is the kernel of the matrix element,
$Q^2=(p_f-p_i)^2$, $P_\mu=(p_f+p_i)_\mu$.
In Refs.[1] the pion form factor at large momentum transfer has been studied by perturbative QCD. 
The perturbative QCD predicts that one gluon exchange dominates $T_H$ and   
at large $Q^2$
\begin{equation} 
F_\pi(Q^2)\sim {\alpha_s(Q^2)\over Q^2}.
\end{equation}
Using a set of pion wave function, in Refs.[1] it is obtained 
\begin{equation}
F_\pi(Q^2)\rightarrow4\pi\alpha_s(Q^2)f^2_\pi/Q^2,
\end{equation}
where the pion decay constant $f_\pi=0.186GeV$ is taken. 

It is known that the wave function of pion is determined by nonperturbative QCD[2].
As a matter of fact, in Ref.[3] three different 
wave functions are chosen to calculate $F_\pi(Q^2)$. Another wave function of pion is discussed 
in Ref.[4].  
In this short note a new wave function of pion is introduced to calculate $F_\pi(Q^2)$ at large
$Q^2$. Eq.(2) is valid at large $Q^2$. The
range of $F_\pi(Q^2)$ in which the behavior(2) is valid is another issue. We try to find the range 
in this note.

A proper wave function of pion determined by nonperturbative QCD 
should at least satisfy two criteria: determining the pion decay
constant $f_\pi$ and the form factor at lower $Q^2$ in both space-like and time-like regions.
Nonperturbative QCD is not solved analytically. It is well known that current algebra is successful
in studying hadron physics at lower energies. The vector and axial-vector currents are constructed
by quark fields. Chiral symmetry is one of the features of QCD. Therefore, current algebra has
close relation with nonperturbative QCD.
In Ref.[5] based on current algebra and chiral symmetry an effective chiral Lagrangian 
of pseudoscalar, vector, and axial-vector mesons is constructed as
\begin{eqnarray}
{\cal L}=\bar{\psi}(x)(i\gamma\cdot\partial+\gamma\cdot v
+\gamma\cdot a\gamma_{5}
-mu(x))\psi(x)-\bar{\psi(x)}M\psi(x)\nonumber \\
+{1\over 2}m^{2}_{0}(\rho^{\mu}_{i}\rho_{\mu i}+
\omega^{\mu}\omega_{\mu}+a^{\mu}_{i}a_{\mu i}+f^{\mu}f_{\mu})
\end{eqnarray}
where \(a_{\mu}=\tau_{i}a^{i}_{\mu}+f_{\mu}\), \(v_{\mu}=\tau_{i}
\rho^{i}_{\mu}+\omega_{\mu}\),
\(u=exp\{i\gamma_{5}(\tau_{i}\pi_{i}+
\eta)\}\), and M is the matrix of current quark masses. 
m is the constituent quark mass and it originates in quark condensation. Therefore,
this theory has dynamical chiral symmetry breaking. 
Integrating out the quark fields, the Lagrangian
of mesons is obtained. The tree diagrams are at leading order in $N_C$ expansion and loop diagrams of 
mesons are at higher order. It is known that quark condensation and $N_C$ expansion are from 
nonperturbative QCD. Therefore, the effects of nonperturbative QCD are embedded in the Lagrangian(4). 
Meson physics has been extensively studied by this Lagrangian. Theory agrees with
data very well. Details can be found in Ref.[5].
By normalizing
the kinetic term of pion to ${1\over2}\partial_\mu\pi^i\partial_\mu\pi^i$,
the pion decay constant is defined 
\begin{equation}
f^2_\pi=F^2(1-{2c\over g}),
\end{equation}
where 
\begin{equation}
{F^2\over16}=\frac{N_C}{(4\pi)^2}{D\over4}\int d^4k\frac{m^2}{(k^2+m^2)^2},
\end{equation}
k is in Euclidean space. 
In this theory the pion fields have two sources: \(u=exp\{i\gamma_{5}\pi\}=
1+i\gamma_{5}\pi+...\) and the shifting $a_\mu\rightarrow a_\mu(physical)-{c\over g}\partial_\mu\pi$. 
The 
shifting is caused by the mixing term $a^i_\mu\partial_\mu\pi^i$. c is determined to be
\begin{equation}
c=\frac{f^2_\pi}{2gm^2_\rho}
\end{equation}
and g is obtained from normalizing the kinetic term of the vector fields and determined to be 0.39 by 
fitting the decay rate of $\rho\rightarrow ee^+$.  
After normalization the pion field of Eq.(4) is redefined as 
${2\over f_\pi}\pi$.
Combining these two sources together, the vertex related to pion is obtained as
\begin{equation}
{\cal L}_\pi=-{2im\over f_\pi}\bar{\psi}\tau_i\gamma_5(1+i{c\over g}{\gamma\cdot\partial\over m})
\psi\pi_i,
\end{equation}
where the differential operator only acts on pion field. It is worth to mention that the second 
term of Eq.(8) is very important in understanding the meson physics related to pion[5]. 
Pion form factor is derived[6]
\begin{eqnarray}
|F_\pi (q^2)|^2 & = & f^2_{\rho \pi \pi }(q^2)
\frac{m_\rho ^4+q^2\Gamma _\rho
^2(q^2)}{(q^2-m_\rho ^2)^2+q^2\Gamma _\rho ^2(q^2)},\nonumber \\
f_{\rho \pi \pi }(q^2)& = & 1+\frac{q^2}{2\pi ^2f_\pi
^2}[(1-\frac{2c}{g})^2-4\pi ^2c^2].
\end{eqnarray}
The radius of charged pion is obtained from Eq.(9)
\begin{equation}
<r^2>_\pi={6\over m^2_\rho}+{3\over \pi^2 f^2_\pi}\{(1-{2c\over g})^2-4\pi^2 c^2\}.
\end{equation}
The second term of Eq.(10) is the new result obtained from this theory[5] and it makes significant 
contribution to $<r^2>_\pi$. Both the form factor and the radius agree with data very well. 
In Refs.[6] the vertices are
derived up to the $4^{th}$ order in derivative and the agreement between the theory and the
experiments in both space-like and time-like regions 
up to $|Q|^2\sim(1.2GeV)^2$.
In Eqs.(9,10) the terms
obtained from the shifting of $a_\mu$ field plays an essential role. Without these terms theoretical 
results do not fit the data. 

Using the vertex(8), the wave function of pion is derived as
\begin{eqnarray}
\phi(z,p)&=&<0|\{\psi({z\over2})\bar{\psi}(-{z\over2})\}|\pi(p)>\nonumber \\
&=&\frac{2\sqrt{2}m}{f_\pi}\frac{1}{(2\pi)^4}
\int d^4k\frac{e^{-ikz}}{(k^2-m^2)(k-p)^2-m^2)}(\gamma\cdot k+m)\gamma_5
(1+{c\over g}{\gamma\cdot p\over m})
(\gamma\cdot k-\gamma\cdot p+m).
\end{eqnarray}
As mentioned above, $f_\pi$ is defined by normalizing the kinetic term of pion fields. On the other hand,
$f_\pi$ can be derived by the pion wave function(11)
\begin{equation}
Tr\phi(0,p)\gamma_\mu\gamma_5={i\over \sqrt{2}}f_\pi p_\mu.
\end{equation}
Substituting Eq.(11) into Eq.(12), Eq.(5) is revealed. 
In the study of pion form factor(9) the Vector Meson Dominance(VMD) is used.
The VMD is a natural result of Eq.(4)[5]. 
It is equivalent to reexpress $F_\pi(Q^2)$ done in Refs.[6] by 
pion wave function(11). Therefore, the pion wave function obtained from Eq.(11) passes the two tests:
determinations of $f_\pi$ and $F_\pi(Q^2)$ at lower values of $Q^2$. 

At large $Q^2$ perturbative QCD is working[1].
Using the pion wave function(11) and $T_H$ with one gluon exchange as given 
in Res.[1], the matrix element of current of pion(1) is written as
\begin{eqnarray}
\lefteqn{<\pi^+|j_\mu|\pi^+>={2m^2\over f^2_\pi}Tr\lambda^a\lambda^a g^2_s\int d^4 k_1 d^4 k_2}\nonumber \\&&
\{\frac{1}{(k_1-k_2
+p_i-p_f)^2}\frac{1}{(k_1+p_i-p_f)^2}Tr\gamma_\nu\phi_\pi(k_1,p_f)\gamma_\mu \gamma\cdot(k_1+p_i-p_2)
\gamma_\nu\phi_\pi(k_2,p_i)\nonumber \\
&&+\frac{1}{(k_1-k_2+p_i-p_f)^2}\frac{1}{(k_2+p_f-p_i)^2}Tr\gamma_\nu\phi_\pi(k_1,p_f)
\gamma_\nu\gamma\cdot
(k_2+p_f-p_i)\gamma_\mu\phi_\pi(k_2,p_i)\},
\end{eqnarray}
where $\phi_\pi(k,p)$ is the pion wave function in momentum space, k is the internal momentum and p is
the momentum of the pion,
\begin{eqnarray} 
\phi_\pi(k_1,p_f)&=&\frac{1}{(2\pi)^4}\frac{1}{(k^2_1-m^2)[(k_1-p_f)^2-m^2]}[\gamma\cdot(k_1-p_f)+m]
\gamma_5(1-{c\over g}{\gamma\cdot p_f\over m})(\gamma\cdot k_1+m),\nonumber\\
\phi_\pi(k_2,p_i)&=&\frac{1}{(2\pi)^4}\frac{1}{(k^2_1-m^2)[(k_2-p_i)^2-m^2]}(\gamma\cdot k_2+m)\gamma_5
(1+{c\over g}{\gamma\cdot p_i\over m})[\gamma\cdot(k_2-p_i)+m].
\end{eqnarray}

In Ref.[5] the universal coupling constant g is defined as
\begin{equation}
g^2={2\over3}\frac{N_C}{(4\pi)^4}\int d^4k\frac{1}{(k^2+m^2)^2}.
\end{equation}
The integral is in Euclidean space. As mentioned above g takes finite value, therefore, a cut-off 
$\Lambda$ of the integral has to be introduced and is determined by the value of g as $\Lambda=1.8GeV$. 
For $Q^2>>(1.8GeV)^2$
Eq.(13) is rewritten as
\begin{eqnarray}
\lefteqn{<\pi^+|j_\mu|\pi^+>={2m^2\over f^2_\pi}Tr\lambda^a\lambda^a g^2_s
{1\over Q^4}\int d^4 k_1 d^4 k_2}\nonumber \\
&&\{
Tr\gamma_\nu\phi_\pi(k_1,p_f)\gamma_\mu \gamma\cdot(k_1+p_i-p_2)
\gamma_\nu\phi_\pi(k_2,p_i)\nonumber \\
&&+Tr\gamma_\nu\phi_\pi(k_1,p_f)\gamma_\nu\gamma\cdot
(k_2+p_f-p_i)\gamma_\mu\phi_\pi(k_2,p_i)\},
\end{eqnarray}
The pion form factor at $Q^2>>(1.8GeV)^2$ is obtained 
\begin{equation}
F_\pi(Q^2)=4\pi\alpha_s(Q^2)f^2_\pi{1\over Q^2}{1\over18}(1-{2c\over g})^{-2}\{{2c^2\over g^2}
+(1-{c\over g})(1-{4c\over g})-{1\over 4\pi^2 g^2}(1-{c\over g})(1-{2c\over g})\}.
\end{equation}
The numerical result is
\begin{equation}
F_\pi(Q^2)=2.65\times10^{-2}4\pi\alpha_s(Q^2)f^2_\pi{1\over Q^2}.
\end{equation}

In summary, a new wave function of pion which is successful in obtaining $f_\pi$ and the pion form factor at 
lower $Q^2$ is used to calculate the pion form factor at very large $Q^2$.
The $F_\pi(Q^2)$ obtained in this short note is much smaller than the one presented in Refs.[1]. 
The Eq.(17) is valid in the range $Q^2>>(1.8GeV)^2$. 

\end{document}